\documentclass[preprint,aps,prd,superscriptaddress,showpacs,nofootinbib]
{revtex4}%
\usepackage{graphicx}
\usepackage{amsmath}
\usepackage{amssymb}
\usepackage{bm}
\usepackage{epsfig}
\def\sl{\!\!\!/}
\begin{document}
\preprint{}
\title{\mbox{}\\[10pt]
Relativistic corrections to Higgs boson decays to quarkonia
}
\author{Geoffrey~T.~Bodwin}
\email[]{gtb@anl.gov}
\affiliation{High Energy Physics Division, Argonne National Laboratory,
Argonne, IL 60439, USA}
\author{Hee~Sok~Chung}
\email[]{chungh@anl.gov}
\affiliation{High Energy Physics Division, Argonne National Laboratory, 
Argonne, IL 60439, USA}
\author{June-Haak~Ee}
\email[]{chodigi@gmail.com}
\affiliation{Department of Physics, Korea University, Seoul 136-713, Korea}
\author{Jungil~Lee}
\email[]{jungil@korea.ac.kr}
\affiliation{Department of Physics, Korea University, Seoul 136-713, Korea}
\author{Frank~Petriello}
\email[]{f-petriello@northwestern.edu}
\affiliation{High Energy Physics Division, Argonne National Laboratory,
Argonne, IL 60439, USA}
\affiliation{Department of Physics, Northwestern University, Evanston, IL 60208, USA}
\date{\today}
\begin{abstract}
We improve the theoretical predictions for the decays of the Higgs boson
to an $S$-wave vector quarkonium plus a photon by calculating the
relativistic correction of order $v^2$, where $v$ is the heavy-quark
velocity in the quarkonium rest frame. Our numerical results are given
for the $J/\psi$ and $\Upsilon(nS)$ channels, with $n=1,2,3$.  The
numerical results include a previously calculated correction of order
$\alpha_s$ and summations, to all orders in $\alpha_s$, of leading
logarithms of $m_H^2/m_Q^2$, where $m_H$ is the Higgs-boson mass and
$m_Q$ is the heavy-quark mass. These QCD corrections apply to the
contribution of leading order in $v$ and to part of the order-$v^2$
correction. For the remainder of the order-$v^2$ correction, we sum
leading logarithms of $m_H/m_Q$ through order $\alpha_s^2$. These
refinements reduce the theoretical uncertainties in the
direct-production amplitudes for $H \to J/\psi+\gamma$ and $H \to
\Upsilon(1S)+\gamma$ by approximately a factor of 3 and open the
door to improved determinations at the LHC of the Higgs-boson Yukawa
couplings to the charm and bottom quarks.
\end{abstract}

\pacs{14.80.Bn, 14.40.Pq}
\vspace{-1cm}
\maketitle

\section{Introduction}
A primary activity of the LHC program is the exploration of the
properties of the Higgs boson, which was discovered over two years
ago by the ATLAS and CMS collaborations~\cite{:2012gk,:2012gu}.
Currently, only couplings to gauge bosons and third-generation fermions
are measured directly~\cite{atlascoup,cmscoup}.  The couplings
that are fixed through the well-measured diboson decays of the
Higgs are determined at the $20\%$--$30\%$ level. 
No deviations from the
predictions of the Standard Model (SM) have been observed.

While the possibility of measuring the Higgs-boson couplings to muons at
the high-luminosity LHC (HL-LHC) has been
studied~\cite{ATLAS-CONF-2013-010,CMS-PAS-HIG-13-007,Dawson:2013bba},
the couplings of the Higgs boson to first- and second-generation quarks
are {\it terra incognita}.  They are only weakly constrained by the
inclusive Higgs-boson production cross sections, yet they can deviate
significantly from their SM values in numerous theories of new physics.
It was long thought to be impossible to measure these couplings, owing
to the severe experimental difficulties that are inherent in
reconstructing the signal and isolating it from the background.

Recent work has demonstrated that there is hope to determine the
Yukawa couplings of first- and second-generation quarks at future runs
of the LHC.  Much of this renewed interest has arisen because of the
realization that exclusive decays of the Higgs boson to vector mesons can
probe its couplings to light quarks.  The resulting final states are
relatively clean experimentally, and the theoretical predictions are
also under control.  The first manifestation of this idea was the
discovery that decays of the Higgs boson to an $S$-wave vector
quarkonium plus a photon ($H\to V+\gamma$) provide opportunities to
determine the $Hc\bar c$ and $Hb\bar b$ couplings
\cite{Bodwin:2013gca}.\footnote{It has
also been realized that decays to light mesons might be used to map out
the structure of Yukawa couplings of the Higgs boson to first- and
second-generation quarks~\cite{Kagan:2014ila}.} [Here, $c$($b$) and
$\bar c$($\bar b$) denote a charm (bottom) quark and charm (bottom)
antiquark.]  While the $Hc\bar c$ coupling might be probed at the LHC by
making use of charm-tagging techniques~\cite{Delaunay:2013pja}, its
phase must be determined through processes that involve quantum
interference effects, such as the decay $H\to J/\psi+\gamma$.

It is our intention in this paper to refine the theoretical prediction
for the $H\to V+\gamma$ processes, where $V=J/\psi$ or $\Upsilon(nS)$,
with $n=1,2,3$.  These modes feature clean experimental signatures in
which a high-transverse-momentum lepton pair recoils against a photon.
They proceed through two distinct mechanisms:
\begin{itemize}

\item In the {\it direct process}, the Higgs boson decays into a
heavy quark-antiquark ($Q\bar Q$) pair, one of which radiates a photon
before forming a quarkonium with the other element of the pair.

\item In the {\it indirect process}, the Higgs boson decays through a
top-quark loop or a vector-boson loop to a $\gamma$ and a $\gamma^*$
(virtual photon). The $\gamma^*$ then decays into a vector quarkonium. 

\end{itemize}
\begin{figure}                                                   
\includegraphics[width=0.8\columnwidth]{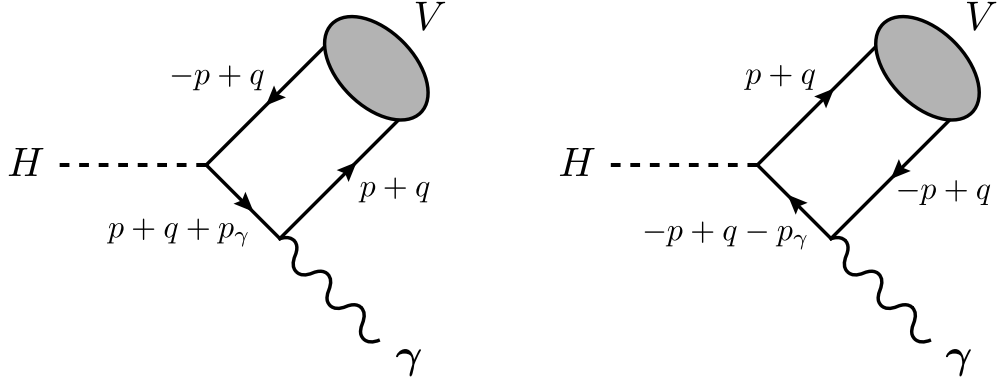}
\caption{\label{figure:direct}
The Feynman diagrams for the direct amplitude for $H\to V+\gamma$ at
order $\alpha_s^0$. The shaded blob represents the quarkonium
wave function. The momenta that are adjacent to the heavy-quark lines
are defined in the text.
}
\end{figure}
\begin{figure}                                                   
\includegraphics[width=0.4\columnwidth]{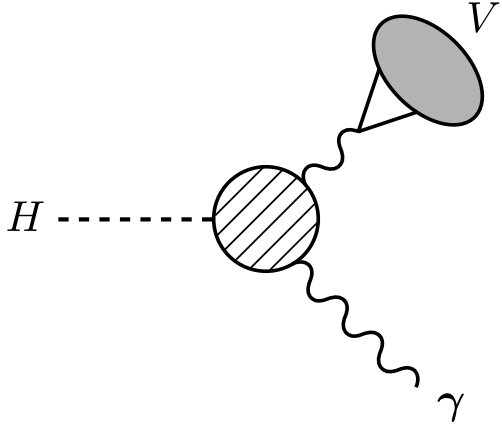}
\caption{\label{figure:indirect}
The Feynman diagram for the indirect amplitude for $H\to V+\gamma$. 
The hatched circle represents top-quark or $W$-boson loops, 
and the shaded blob represents the quarkonium wave function.
}
\end{figure}
The Feynman diagrams for the direct and indirect processes are
shown in Figs.~\ref{figure:direct} and \ref{figure:indirect},
respectively.
It is the quantum interference between these two processes that provides
phase information about the $Hc\bar c$ and $Hb\bar b$ couplings. The
interference is destructive. In the case of the decay to the 
$\Upsilon$, the destructive interference is nearly complete, and so the 
rate is very sensitive to the $Hb\bar b$ coupling.

The indirect decay amplitudes are determined at percent-level accuracy.
The partial amplitude for the Higgs-boson decay to $\gamma\gamma^*$ can
be inferred from calculations of the $H \to \gamma\gamma$ rate
\cite{Dittmaier:2011ti,Dittmaier:2012vm}. The coupling of the quarkonium
to a virtual photon is known from the decay rate of the quarkonium to a
lepton pair.

The largest theoretical uncertainty in the direct amplitude for $H\to
J/\psi+\gamma$ and, consequently, in the decay rate, arises from
uncalculated relativistic corrections. These corrections take into
account the relative motion of the $Q$ and $\bar Q$  in the quarkonium.
They are nominally of order $v^2$, where $v$ is the rms velocity of the $Q$
or $\bar Q$ in the quarkonium rest frame. $v^2\approx 25\%$ for the
$J/\psi$ and $v^2\approx 10\%$ for the $\Upsilon$.

In this paper, we compute 
order-$v^2$ corrections and some order-$\alpha_s v^2$ corrections to 
the direct
amplitudes for the processes $H\to J/\psi+\gamma$ and $H\to
\Upsilon(nS)+\gamma$, where $\alpha_s$ is the strong coupling. We also include
some corrections involving leading logarithms of $m_H^2/m_Q^2$ that are
of order $v^2$ and of higher orders in $\alpha_s$. (Here, $m_H$ is the
Higgs-boson mass and $m_Q$ is the heavy-quark mass.)

The remainder of this paper is organized as follows: In
Sec.~\ref{sec:NRQCD}, we use the methods of nonrelativistic QCD
(NRQCD) factorization~\cite{Bodwin:1994jh} to compute the relativistic
corrections to $H\to V+\gamma$. These corrections can also be computed,
in the limit $m_V/m_H\to 0$, where $m_V$ is the quarkonium mass, by
making use of light-cone methods~\cite{Lepage:1980fj,Chernyak:1983ej}.
We carry out the light-cone calculation of the relativistic corrections
in Sec.~\ref{sec:light-cone}. The light-cone computation allows us to
take advantage of existing calculations of corrections of
next-to-leading order in $\alpha_s$ and is a convenient framework in
which to compute logarithms of $m_H^2/m_Q^2$. We give numerical results
for the decay rates in Sec.~\ref{sec:numerical} and summarize our
findings in Sec.~\ref{sec:summary}.

\section{NRQCD calculation \label{sec:NRQCD}}
In this section we compute relativistic corrections to the direct
amplitude for $H\to V+\gamma$ by making use of the standard methods of
NRQCD factorization \cite{Bodwin:1994jh}. We begin by considering the
amplitude for $H\to Q\bar Q+\gamma$, where the $Q\bar Q$ pair is in a
color-singlet, spin-triplet $S$-wave state. We take the Higgs-boson,
$Q$, $\bar Q$, and $\gamma$ momenta to be $p_H$, $p_1=p+q$, $p_2=p-q$,
and $p_\gamma$, respectively. These momenta satisfy the following
relations: 
\begin{eqnarray}
p_H&=&2p+p_\gamma,\;\;\;\;
p\cdot q=0,\;\;\;\;
p_H^2=m_H^2,\nonumber \\
p_1^2&=&m_Q^2,\;\;\;\;\;\;\;\;\;\;
p_2^2=m_Q^2,\;\;\;\;
p_\gamma^2=0,\nonumber \\
p^2&=&E^2,\;\;\;\;\;\;\;\;\;\;\;
E^2\equiv m_Q^2-q^2\equiv m_Q^2(1+v^2). 
\end{eqnarray}
In the $Q\bar Q$ rest frame, $p=(E,\bm{0})$ and $q=(0,\bm{q})$.

We take the polarization of the $\gamma$ to be $\epsilon_\gamma$, 
and we take the spin polarization of the $Q\bar Q$ pair to be
$\epsilon(\lambda)$, where $\lambda$ is the polarization state. The
color-singlet, spin-triplet projector, correct to all orders in $v$, is
given by~\cite{Bodwin:2002hg}
\begin{equation}
\label{Pi3}
\Pi_3(p_1, p_2, \lambda)
=
\frac{1}{8 \sqrt{2} E^2 (E+m_Q)} 
(p\sl_2 -m_Q) 
\,\epsilon\sl^* (\lambda) (p\sl_1 + p\sl_2 + 2 E) (p\sl_1+m_Q)
\otimes \frac{{\bf 1}}{\sqrt{N_c}},
\end{equation}
where $\bf 1$ is the unit color matrix and $N_c=3$ is the
number of colors.

The $H\to Q\bar Q +\gamma$ amplitude arises from two Feynman diagrams,
which are shown in Fig.~\ref{figure:direct}. For a color-singlet,
spin-triplet $Q\bar Q$ pair, it is given by
\begin{eqnarray}
i {\cal M}_{\rm dir}[Q\bar Q({\rm triplet})] &=& 
-ie e_Q\kappa_Q m_Q (\sqrt{2} G_F)^{\frac{1}{2}}
{\rm Tr} \bigg\{
\bigg[\frac{(-p\sl+q\sl-p\sl_\gamma+m_Q)\,\epsilon\sl^*_\gamma}{
(p-q+p_\gamma)^2-m_Q^2+i \varepsilon} 
\nonumber \\ && + 
\frac{\,\epsilon\sl^*_\gamma (p\sl+q\sl+p\sl_\gamma+m_Q)}{
(p+q+p_\gamma)^2-m_Q^2+i \varepsilon} \bigg]
 \Pi_3(p+q,p-q,\lambda) 
\bigg\},
\label{trip-amplitude}
\end{eqnarray}
where the trace is over the gamma and the color matrices, $e$ is the
electromagnetic coupling, $G_F$ is the Fermi weak coupling, $e_Q$ is the
fractional heavy-quark charge, and $\kappa_Q$ is an adjustable factor in
the $HQ\bar Q$ coupling. $\kappa_Q=1$ in the SM.

Owing to charge-conjugation symmetry, the two contributions in
Eq.~(\ref{trip-amplitude}) differ only by a change of sign of $q$. We
obtain the $S$-wave contribution by averaging over the angles of
$\bm{q}$ in the $Q\bar Q$ rest frame. In that average, contributions
that are odd in $q$ vanish. Hence, we can write the spin-triplet,
$S$-wave amplitude as
\begin{equation}
i {\cal M}[Q\bar Q(^3S_1)]=
-2 i e e_Q \kappa_Q m_Q \big(\sqrt{2} G_F\big)^{\!\frac{1}{2}}
{\int}_{\!\!\!\!\hat{\bm{q}}} 
{\rm Tr} \bigg[ \Pi_3(p+q,p-q,\lambda) 
\frac{(-p\sl+q\sl-p\sl_\gamma+m_Q)\epsilon\sl_\gamma^*}{
(p-q+p_\gamma)^2-m_Q^2+i \varepsilon} 
\bigg], 
\label{3S1-amplitude}
\end{equation}
where a factor of 2 takes into account both contributions in
Eq.~(\ref{trip-amplitude}) and the symbol ${\int}_{\!\!\hat{\bm{q}}}$
denotes the average over the direction of $\hat{\bm{q}}\equiv
\bm{q}/|\bm{q}|$ in the rest frame of $V$:
\begin{equation}
{\int}_{\!\!\!\!\hat{\bm{q}}} 
\equiv\int\!\frac{d \Omega_{\hat{\bm{q}}}}{4 \pi}. 
\end{equation} 

Evaluation of the trace in Eq.~(\ref{3S1-amplitude}) gives 
\begin{eqnarray}                                                   
i {\cal M}_{\rm dir}[Q\bar Q(^3S_1)]&=&
-2 i e e_Q \kappa_Q m_Q (\sqrt{2} G_F)^{1/2}                             
{\int}_{\!\!\!\!\hat{\bm{q}}}\,
\frac{-\sqrt{N_c}}{2\sqrt{2} E^2[(p-q+p_\gamma)^2-m_Q^2+i \varepsilon]}
\nonumber\\
&&\times\bigg[ \frac{(m_H^2 + 4 E^2 + 8 E m_Q )}{E+m_Q}\, 
\epsilon^{*}_\gamma \cdot q \,\epsilon^{*} \cdot q 
- \frac{4p_\gamma \cdot q }{E+m_Q}\, 
\epsilon_\gamma^{*} \cdot p\,\epsilon^{*} \cdot q 
\nonumber \\ 
&&\qquad - 8 E \epsilon^{*}_\gamma \cdot p \,\epsilon^{*} \cdot q 
+ 4 m_Q \epsilon^{*}_\gamma \cdot p \,\epsilon^{*} \cdot p_\gamma 
- (m_H^2-4 E^2 ) m_Q \epsilon^{*}_\gamma \cdot \epsilon^{*} \bigg]. 
\end{eqnarray}

We can write the quark-propagator denominator as $2(p-q)\cdot
p_\gamma$. Then, the amplitude in Eq.~(\ref{3S1-amplitude}) contains the
tensor integrals
\begin{subequations}
\begin{eqnarray}
I &=& {\int}_{\!\!\!\!\hat{\bm{q}}}\,
\frac{p\cdot p_\gamma}{(p-q)\cdot p_\gamma},
\\
I^\mu &=&  {\int}_{\!\!\!\!\hat{\bm{q}}}\,
\frac{p\cdot p_\gamma}{(p-q)\cdot p_\gamma}\,q^\mu,
\\
I^{\mu \nu} &=&  {\int}_{\!\!\!\!\hat{\bm{q}}}\,
\frac{p\cdot p_\gamma}{(p-q)\cdot p_\gamma}\,q^\mu q^\nu.
\end{eqnarray}
\end{subequations}
Because $q\cdot p=0$, the tensor integrals $I^\mu$ and $I^{\mu\nu}$
must be orthogonal to $p$: $I^\mu p_\mu=0$, $I^{\mu\nu}p_\mu=I^{\mu\nu}p_\nu=0$.
Therefore, it is convenient to define the four-vector
\begin{equation}
\bar p_\gamma \equiv p_\gamma - \frac{p_\gamma \cdot p}{p^2} p, 
\end{equation}
which is orthogonal to $p$. 
From the orthogonality of $I^\mu$ and $I^{\mu\nu}$ to $p$, it follows
that $I^\mu$ must be proportional to $\bar{p}_\gamma^\mu$ and that
$I^{\mu\nu}$ must be a linear combination of $-g^{\mu\nu}+({p^\mu
p^\nu})/{p^2}$ and $\bar{p}_\gamma^\mu\bar{p}_\gamma^\nu$. A
straightforward analysis then shows that
\begin{subequations}
\begin{eqnarray}
I&=&  L(\delta)\equiv
\frac{1}{2 \delta} \log \frac{1+\delta}{1-\delta},\\
I^\mu &=& \frac{4E^2(1-I)}{m_H^2 - 4E^2} \,\bar p^\mu_\gamma ,\\
I^{\mu \nu}&=& 
\frac{E^2-m_Q^2 I}{2}
\left(-g^{\mu \nu}+\frac{p^\mu p^\nu}{p^2}\right)
+\frac{8E^2 [ (m_Q^2 + 2E^2)I - 3 E^2]}{(m_H^2 -4E^2)^2} 
\,\bar p^\mu_\gamma \bar p^\nu_\gamma,
\end{eqnarray}
where 
\begin{equation}
\delta=\frac{\sqrt{-q^2}}{E}=\frac{|\bm{q}|}{E}=\frac{v}{\sqrt{1+v^2}}.
\end{equation}
\end{subequations}

Now the amplitude can be written as 
\begin{subequations}
\begin{equation}
\label{Mdir3s1}
i {\cal M}_{\rm dir}[Q\bar Q(^3S_1)]=i {\cal M}_{\rm dir}^{(0)}
[Q\bar Q(^3S_1)] R(v^2),
\end{equation}
where
\begin{equation}
\label{M0dir3s1}
i {\cal M}_{\rm dir}^{(0)}[Q\bar Q(^3S_1)]=
ie e_Q \kappa_Q 
\big(\sqrt{2}G_F\big)^{\!\frac{1}{2}}\sqrt{2N_c} 
\left(
-\epsilon^*\cdot \epsilon^*_{\gamma}
+\frac{\epsilon^*\cdot p_\gamma  \,
p\cdot\epsilon^*_{\gamma}}
{p_\gamma\cdot p}
\right),
\end{equation}
is the amplitude in order $v^0$, and the factor $R(v^2)$, which 
contains the relativistic corrections, is given by
\begin{equation}
R(v^2)=
\frac{m_Q}{2E^2}
\left\{
\frac{E^2+m_Q(2E+m_Q)L(\delta)}{E+m_Q}
+\frac{8E[E^2-m_Q^2L(\delta)]}{m_H^2-4E^2}
\right\}.
\label{R-v2}
\end{equation}
\end{subequations}
The invariance under electromagnetic gauge transformations is manifest
in the last factor in Eq.~(\ref{M0dir3s1}). In a physical gauge in the
$H$ rest frame, $p\cdot \epsilon_\gamma=0$, and the last term in the last
factor in Eq.~(\ref{M0dir3s1}) vanishes. Hence, the expression in
Eq.~(\ref{M0dir3s1}) is independent of $v$.\footnote{One can also see
that the expression (\ref{M0dir3s1}) is independent of $v$ from the fact
that the $v$ dependence of the four-vector $p$ is contained in a factor
that is common to all of the components of $p$. That factor cancels in the
expression (\ref{M0dir3s1}).} 

Now we can obtain the physical amplitude by carrying out the standard
matching procedure between NRQCD and full QCD \cite{Bodwin:1994jh}. That
is, we write $i {\cal M}_{\rm dir}$ in terms of NRQCD long-distance
matrix elements (LDMEs) and determine the corresponding short-distance
coefficients by comparing the NRQCD expression, evaluated in the $Q\bar
Q(^3S_1)$ state, with Eq.~(\ref{Mdir3s1}). Having determined the
short-distance coefficients, we obtain the physical amplitude by
evaluating the NRQCD LDMEs in the physical quarkonium state. We
find that the direct amplitude for $H\to V+\gamma$ is given by
\begin{subequations}
\begin{equation}
i {\cal M}_{\rm dir}
[H \to V+\gamma]=\sqrt{2m_V}\phi_0\,
i\mathcal{M}_{\rm dir}^{(0)}[H \to V+\gamma]
\sum_{n=0}^\infty\frac{1}{n!}
\left. \left( \frac{\partial }{\partial v^2} \right)^{\!n} \!R(v^2)
\right|_{v=0} \langle 
v^{2n}\rangle,
\label{phys-amp}
\end{equation}
where 
\begin{equation}
i {\cal M}_{\rm dir}^{(0)}
[H\to V+\gamma]\equiv
ie e_Q \kappa_Q (\sqrt{2} G_F)^{\,\frac{1}{2}} \sqrt{2N_c}
\left(-\epsilon_V^*\cdot \epsilon^*_{\gamma}
+\frac{\epsilon_V^*\cdot p_\gamma  \,
p_V\cdot\epsilon^*_{\gamma}}
{p_\gamma\cdot p_V}\right),
\label{amp-v0}
\end{equation}
and $p_V$, $m_V$, and $\epsilon_V$ are the momentum, mass, and
polarization of the quarkonium.\footnote{Owing to the denominator
factors $p$ and $p_V$ in the expressions in Eqs.~(\ref{M0dir3s1}) and
(\ref{amp-v0}), the corresponding NRQCD LDMEs contain nonlocal
operators. One can avoid the appearance of these nonlocal operators in
the matching procedure by working in a physical gauge, in which $p\cdot
\epsilon_\gamma=p_V\cdot\epsilon_\gamma=0$, so that the second term in
parentheses in Eqs.~(\ref{M0dir3s1}) and (\ref{amp-v0}) vanishes. These
terms can then be restored by requiring the final expression to be
manifestly gauge invariant.} The quantity $\langle v^{2n}\rangle$ is
given by a ratio of NRQCD LDMEs:
\begin{equation}
\langle v^{2n}\rangle=
\frac{1}{m_Q^{2n}} 
\frac{\langle V(\bm{\epsilon})
|\psi^\dagger(-\frac{i}{2} 
\tensor{\bm{\nabla}})^{2n}\bm{\sigma}\cdot\bm{\epsilon}\chi|0\rangle}
{\langle V(\bm{\epsilon})
|\psi^\dagger \bm{\sigma}\cdot\bm{\epsilon}\chi|0\rangle}.
\label{v2n-def}
\end{equation}
$\phi_0$ is the quarkonium wave function at the origin, which is given by
\begin{equation}
\phi_0=\frac{1}{\sqrt{2N_c}}
\langle V(\bm{\epsilon})
|\psi^\dagger \bm{\sigma}\cdot\bm{\epsilon} \chi|0\rangle.
\end{equation}
\end{subequations}
In the LDMEs, $\psi$ is the two-component (Pauli) spinor field that
annihilates a heavy quark, and $\chi$ is the two-component spinor field
that annihilates a heavy antiquark. The factor $\sqrt{2m_V}$ in
Eq.~(\ref{phys-amp}) arises from the relativistic normalization of the
quarkonium state. In this factor and in the phase space, we choose $m_V$
to be the physical quarkonium mass, rather than the mass of the $Q\bar
Q$ state ($2E$).

In Eq.~(\ref{phys-amp}), we have neglected contributions from LDMEs that
involve factors of the gauge field. These contributions first appear in
order $v^4$. In this paper, we work through order $v^2$. Retaining only
contributions through order $v^2$ in Eq.~(\ref{phys-amp}), we obtain
\begin{eqnarray}
i {\cal M}_{\rm dir} [H \to V+\gamma]&=&\sqrt{2m_V}\phi_0\,i
{\cal M}_{\rm dir}^{(0)} [H \to V+\gamma] 
\left[1 - \frac{3 m_H^2 - 28 m_Q^2}{6 (m_H^2 -4 m_Q^2)} \langle 
v^2\rangle+ O(\langle v^4\rangle) \right] \nonumber\\
&\approx&\sqrt{2m_V}\phi_0\,i {\cal M}_{\rm dir}^{(0)} 
[H \to V+\gamma]                                          
\left[1 - \frac{1}{2} \langle
v^2\rangle + O(\langle v^4\rangle)\right], 
\label{phys-amp-v2}
\end{eqnarray}
where we have dropped contributions of higher order in $m_Q^2/m_H^2$ in
the last line. Our result for the order-$v^0$ amplitude in
Eq.~(\ref{phys-amp-v2}) agrees with those in
Refs.~\cite{Keung:1983ac,Bodwin:2013gca}.

We can assess the convergence of the $v$ expansion for the class of LDMEs in 
Eq.~(\ref{phys-amp}) by making use of the generalized Gremm-Kapustin 
relation \cite{Bodwin:2006dn}
\begin{equation}
\langle v^{2n}\rangle=\langle v^2\rangle^n,
\end{equation}
which holds for dimensionally regulated LDMEs up to corrections of
relative order $v^2$. Taking $\langle v^2\rangle=0.20$, which is the
approximate value for the $J/\psi$,\footnote{Note that the ratio of
LDMEs $\langle v^2\rangle$ is different from the quantity $v^2$ that was
mentioned earlier. $v^2$ is the average of $\bm{q}^2/m_Q^2$ over the
square of the quarkonium wave function in the quarkonium rest frame.
$\langle v^2\rangle$ can be significantly different from $v^2$, in part
because the numerator LDME of $\langle v^2\rangle$ contains a linear
ultraviolet divergence that is subtracted in dimensional regularization.}
we find that the full expression in Eq.~(\ref{phys-amp}) gives a
relativistic correction of $-8.8\%$, while the order-$v^2$ expression in
Eq.~(\ref{phys-amp-v2}) gives a relativistic correction of $-10\%$. The
difference between these corrections, $1.2\%$, is smaller than the
nominal relative size of an order-$v^4$ correction, indicating that the
$v$ expansion is converging well. 
In fact, from the analytic structure of $R(v^2)$, we can see that the radius 
of convergence of the series in $v^2$ is unity.
\section{Light-cone calculation \label{sec:light-cone}}
One can also compute the direct amplitude $i{\cal M}_{\rm
dir}[H\to V+\gamma]$ in the light-cone approach. In leading twist, the
computation is accurate up to corrections of order $m_Q^2/m_H^2$. Our
motivation for examining the light-cone approach is two-fold: (1) we
wish to make contact with the order-$\alpha_s$ light-cone calculation of
$i{\cal M}_{\rm dir}[H\to V+\gamma]$ in Ref.~\cite{Shifman:1980dk};
(2) the light-cone formalism is a convenient one in which to compute
logarithms of $m_H^2/m_Q^2$.
\subsection{Light-cone direct amplitude}
Let us now derive the light-cone amplitude for the direct process at
order $\alpha_s^0$ and at leading twist, that is, at leading order in
$1/m_H$. We work implicitly in the $H$ rest frame and neglect $m_Q$ in
comparison with $m_H$. Hence, the quarkonium momentum $2p$ is lightlike, 
and we take $p$ to be in
the minus light-cone direction. The $H\to V+\gamma$ amplitude for the
direct process is
\begin{eqnarray}
i{\cal M}_{\rm dir}^{\rm LC}[H\to V+\gamma]&=& 
- i e e_Q \kappa_Q m_Q 
\big(\sqrt{2}G_F\big)^{\!\frac{1}{2}}
\int\!\frac{d^4q}{(2\pi)^4}\, 
\langle V | \bar Q(p+q) \nonumber\\
&&\times \bigg[ 
\frac{-p\sl+q\sl-p\sl_\gamma}{(p-q+p_\gamma)^2+i \varepsilon} 
\,/\!\!\!\epsilon^*_\gamma 
+ 
\,/\!\!\!\epsilon^*_\gamma 
\frac{p\sl+q\sl+p\sl_\gamma}{(p+q+p_\gamma)^2+i \varepsilon} 
\bigg]
Q(p-q)| 0 \rangle, 
\end{eqnarray}
where we have set $m_Q=0$, except in the $HQ\bar Q$ coupling.
It is understood that the integration over the 
transverse components of $q$ is dimensionally regulated. The scale of 
the dimensional regularization ultimately sets the scale of the 
light-cone distribution amplitude (LCDA).

Using 
$(/\!\!\!p- q\sl) Q(p-q)=\bar{Q}(p+q)(/\!\!\!p+ q\sl)=0$,
we obtain
\begin{eqnarray}
i {\cal M}_{\rm dir}^{\rm LC} [H\to V+\gamma]
&=& 
- \frac{ie e_Q \kappa_Q m_Q \big(\sqrt{2}G_F\big)^{\frac{1}{2}} }{p_\gamma\cdot p_V}
\!\int\! \frac{d^4q}{(2\pi)^4}\, \langle V | \bar Q(p+q) 
\bigg(
\frac{-p\sl_\gamma \epsilon\sl^*_\gamma}{1-x} 
+ 
\frac{\epsilon\sl^*_\gamma p\sl_\gamma}{1+x} 
\bigg)
\nonumber\\&&\times
Q(p-q)| 0 \rangle\nonumber\\
&=& 
- ie e_Q \kappa_Q m_Q \big(\sqrt{2}G_F\big)^{\frac{1}{2}} 
\frac{\epsilon_\gamma^{*\mu}p_\gamma^\nu}{p_\gamma\cdot p_V}
\!\int\!\frac{d^4q}{(2\pi)^4}\, \langle V | \bar Q(p+q) 
\frac{[\gamma_\mu,\gamma_\nu]}{1-x^2} 
Q(p-q)| 0 \rangle.\nonumber\\
\end{eqnarray}
Here, we have followed the light-cone effective-field-theory procedure. 
That is, we have set $q=xp$, neglecting $q^+$ and $\bm{q}_\perp$,  
in the expression between $\bar Q$ and 
$Q$, which is proportional to the hard-scattering amplitude. However, we have 
retained $q^+$ and $\bm{q}_\perp$ nonzero in the other factors, which 
are proportional to the quarkonium wave function. In the last line,
we have used the fact that $\epsilon_\gamma^*\cdot p_\gamma=0$.

The LCDA $\phi(x)$ is defined by
\begin{equation}
\frac{1}{2}\langle V | \bar Q (z) [\gamma^\mu,\gamma^\nu] 
[z,-z]Q (-z) | 0 \rangle
= f_V (\epsilon_V^*{}^\mu p_V^\nu - \epsilon_V^*{}^\nu p_V^\mu) 
\int_{-1}^{+1}\! dx\, e^{ip^- zx} \phi(x), 
\label{lcda-def}
\end{equation}
where $z$ lies along the plus light-cone direction.
The gauge link $[z,-z]$, which makes the nonlocal operator gauge 
invariant, is given by
\begin{equation}
[z,-z]=P\exp\left[ig_{s}\int_{-z}^{+z}\! dx\, A^+(x)\right], 
\end{equation}
where $g_s=\sqrt{4\pi\alpha_s}$, $A^\mu=A^\mu_a T^a$ is a
matrix-valued gluon field $A^\mu_a$ with the color index $a=1,$ 2,
$\ldots$, $N_c^2-1$; $T^a$ is the generator of the fundamental
representation of SU(3) color; and $P$ denotes path ordering. The gauge
link vanishes in our case because we are working at order
$\alpha_s^0$. (More generally, the gauge link vanishes in the
light-cone gauge $A^+=0$.) It follows from the definition
(\ref{lcda-def}) that 
\begin{equation}
i {\cal M}_{\rm dir}^{\rm LC} [H\to V+\gamma]
= \frac{i}{2}e e_Q \kappa_Q m_Q \big(\sqrt{2}G_F\big)^{\!\frac{1}{2}}
f_V \left(
-\epsilon_{V}^*\cdot \epsilon^*_{\gamma}
+\frac{\epsilon_{V}^*\cdot p_\gamma  \,
p\cdot\epsilon^*_{\gamma}}
{p_\gamma\cdot p}
\right)
\int_{-1}^{+1}\! dx\, T_0(x)\phi(x),
\label{amp-lcda}
\end{equation}
where
\begin{equation}
\label{eq:hard_kernel}
T_0(x)=\frac{4}{1-x^2}
= 4(1+x^2+\ldots)
\end{equation}
is the hard-scattering kernel at leading order in $\alpha_s$.
The result in Eq.~(\ref{amp-lcda}) agrees with the corresponding
expression in Ref.~\cite{Jia:2008ep}.
\subsection{Decay constant $\bm{f_V}$}
Next, we wish to determine the decay constant $f_V$. Setting $z=0$ in 
Eq.~(\ref{lcda-def}) and imposing the normalization condition
\begin{equation}
\int_{-1}^{+1} \!dx\, \phi(x)=1,
\end{equation}
we obtain
\begin{equation}
\langle V | \bar Q[\gamma^\mu,\gamma^\nu]Q| 0 \rangle
= 2 f_V (\epsilon_V^*{}^\mu p_V^\nu - \epsilon_V^*{}^\nu p_V^\mu).
\label{lcda-norm}
\end{equation}
We can evaluate the matrix element on the left side of
Eq.~(\ref{lcda-norm}) in terms of NRQCD LDMEs by making use of the
procedure that we followed in Sec.~\ref{sec:NRQCD}. The result is
\begin{eqnarray}
\langle Q \bar Q({}^3S_1) | \bar Q [ \gamma^\mu, \gamma^\nu] Q | 0 \rangle
&=& 
\int_{\!\hat{\bm{q}}}
{\rm Tr} \left[ \Pi_3 (p+q,p-q,\lambda) ( \gamma^\mu \gamma^\nu
-\gamma^\nu \gamma^\mu) \right]
\nonumber \\
&=& 
\frac{2\sqrt{2 N_c} (E+ 2 m_Q)}{3 E^2} 
(\epsilon^{\mu*} p^\nu - \epsilon^{\nu*} p^\mu) 
\nonumber \\
&=& 2F(v^2) \frac{\sqrt{2N_c}}{m_Q} 
(\epsilon^{\mu*} p^\nu - \epsilon^{\nu*} p^\mu), 
\end{eqnarray}
where
\begin{equation}
F(v^2) = \frac{ m_Q ( E+2 m_Q )}{3 E^2} 
= \frac{2+ \sqrt{1+v^2}}{3 (1+v^2)} 
= 1- \frac{5}{6} v^2 +O(v^4). 
\label{F-v2}
\end{equation}

Then, carrying out the NRQCD matching procedure, we obtain
\begin{equation}
\langle V | \bar Q [ \gamma^\mu, \gamma^\nu] Q | 0 \rangle
= \frac{\sqrt{2N_c} \sqrt{2 m_V}}{m_Q} \phi_0 
(\epsilon_V^{\mu*} p_V^\nu - \epsilon_V^{\nu*} p_V^\mu) 
\sum_{n=0}^\infty
 \frac{1}{n!} \left( \frac{\partial}{\partial v^2} \right)^n 
F(v^2) \bigg|_{v^2=0} \langle v^{2n} \rangle.
\end{equation}
Inserting this result into Eq.~(\ref{lcda-norm}), we find that
\begin{equation}
f_V = \frac{\sqrt{2N_c} \sqrt{2 m_V}}{2m_Q} \phi_0 
{}\sum_{n=0}^\infty
 \frac{1}{n!} \left( \frac{\partial}{\partial v^2} \right)^n 
F(v^2) \bigg|_{v^2=0} \langle v^{2n} \rangle. 
\end{equation}
Hence, from Eq.~(\ref{amp-lcda}), we see that
\begin{eqnarray}
i {\cal M}_{\rm dir}^{\rm LC}
[H\to V+\gamma] &=& 
\sqrt{2 m_V} \phi_0\,
i {\cal M}_{\rm dir}^{(0)}[H\to V+\gamma] 
\sum_{n=0}^\infty
\frac{1}{n!} \left( \frac{\partial}{\partial v^2} \right)^n 
F(v^2) \bigg|_{v=0}
\langle v^{2n} \rangle\nonumber\\
&&\times 
\int_{-1}^{+1}\! dx\, \frac{T_0(x)}{4} \phi(x)\nonumber\\
&=&
\sqrt{2 m_V} \phi_0\,
i {\cal M}_{\rm dir}^{(0)}[H\to V+\gamma]\left[1-\frac{5}{6}\langle v^2\rangle 
+O(\langle v^4\rangle)\right] \nonumber\\
&&\times
\int_{-1}^{+1}\! dx\, \frac{T_0(x)}{4} \phi(x).
\label{phys-amp-lcda}
\end{eqnarray}
\subsection{Relativistic corrections}
Some of the relativistic corrections in the direct
amplitude for $H\to V+\gamma$ are apparent in the factor $F(v^2)$ in
Eq.~(\ref{phys-amp-lcda}). There are additional relativistic corrections
that come from the integral over $x$ in Eq.~(\ref{phys-amp-lcda}). We
make them manifest by carrying out a formal expansion of $\phi(x)$
about $x=0$:
\begin{equation}
\phi(x) = 
\sum_{k=0}^\infty
\frac{(-1)^k \langle x^{k} \rangle}{k!}  \delta^{(k)}(x),
\label{delta-expn}
\end{equation}
where $\delta^{(k)}(x)$ is the $k$th derivative of
the Dirac delta function.
Then,
using the fact that $\phi(x)$ is an even function of $x$, we find that
\begin{eqnarray}
\int_{-1}^{+1} dx\, T_0(x)\phi(x)&=&
4\sum_{k=0}^\infty\int_{-1}^{+1} \!\!\!dx\,x^{2k}
\phi(x)=
4\sum_{k=0}^\infty\langle x^{2k}\rangle
\nonumber\\
&=&4+\frac{4}{3}\langle v^2\rangle + O(\langle v^4\rangle),
\label{TH-expand}
\end{eqnarray}
where 
\begin{equation}
\langle x^n\rangle=\int_{-1}^{+1}\! dx\, x^n\phi(x),
\end{equation}
and we have used the relation \cite{Braguta:2006wr,Braguta:2007fh}
\begin{equation}
\langle x^2\rangle=\frac{1}{3}\langle v^2\rangle,
\label{x2-expect}
\end{equation}
which holds for $S$-wave quarkonia, up to corrections of order 
$\langle v^4\rangle$. 
Then, from Eqs.~(\ref{F-v2}), (\ref{phys-amp-lcda}), and
(\ref{x2-expect}), we have
\begin{equation}
i {\cal M}_{\rm dir}^{\rm LC} [H\to V+\gamma]=
 \sqrt{2 m_V} \phi_0 \, 
i {\cal M}_{\rm dir}^{(0)}[H\to 
V+\gamma]
\Big[1-\frac{1}{2}\langle v^2\rangle
+O(\langle v^4\rangle)\Big],
\end{equation}
in agreement with the last line of Eq.~(\ref{phys-amp-v2}).
\subsection{Evolution of the LCDA}
The LCDA depends on a scale $\mu$. If we employ dimensional
regularization to define and renormalize the LCDA, then $\mu$ is the
scale that is associated with the dimensional regularization. The
evolution with respect to $\mu$ is governed by the evolution equation
\cite{Lepage:1980fj}
\begin{equation}
\mu^2 \frac{\partial}{\partial \mu^2} 
\phi(x,\mu) =  C_F
\frac{\alpha_s(\mu) }{4\pi} 
\int_{-1}^1\! dy\, V_T (x,y) \phi(y,\mu), 
\label{LCDA-evol}
\end{equation}
where 
\begin{subequations}
\begin{eqnarray}
V_T(x,y) &=& V_0 (x,y) 
- \frac{1-x}{1-y} \theta (x-y)
- \frac{1+x}{1+y} \theta(y-x),
\\
V_0(x,y) &=& V_{\rm BL} (x,y) 
- \delta(x-y) \int_{-1}^1\! dz\, V_{\rm BL} (z,x),
\\
V_{\rm BL} (x,y)
&=& \frac{1-x}{1-y} \left( 1+ \frac{2}{x-y} \right) \theta (x-y)
+ \frac{1+x}{1+y} \left( 1+ \frac{2}{y-x} \right) \theta(y-x).
\end{eqnarray}
\end{subequations}

The evolution equation (\ref{LCDA-evol}) is usually solved by
expanding $\phi(x)$ in eigenfunctions of the evolution kernel $V_T$. That
approach is discussed in the Appendix. It was used in
Ref.~\cite{Shifman:1980dk} to obtain a summation of the leading
logarithms of $m_H^2/m_Q^2$ to all orders in $\alpha_s$ for the $x^0$ term
in $T_0$ [Eq.~(\ref{eq:hard_kernel})]. As is explained in the Appendix, this
approach fails to give a convergent expression for physical values of
$m_H$ and $m_Q$ for the $x^2$ term in $T_0$. Therefore, we compute the
logarithms of $m_H^2/m_Q^2$ for the $x^2$ term in $T_0$ by solving the
evolution equation perturbatively.

The solution of Eq.~(\ref{LCDA-evol}) through order $\alpha_s^2$ is 
given by \cite{Jia:2008ep} 
\begin{eqnarray}
\phi(x,\mu) 
&=& \phi(x,\mu_0)
+
\Bigg[C_F
\frac{\alpha_s(\mu)}{4\pi}\log\frac{\mu^{2}}{\mu_0^2}\Bigg]
\Bigg[
1
+\frac{\beta_0}{2}\frac{\alpha_s(\mu)}{4\pi}
\log\frac{\mu^{2}}{\mu_0^2}
\Bigg]
\nonumber\\
&&\times \int_{-1}^1 \!\!dy 
\,\, V_T (x,y) \phi(y,\mu_0)
+
\frac{1}{2}
\Bigg[C_F\frac{\alpha_s(\mu)}{4\pi}
\log\frac{\mu^{2}}{\mu_0^2}\Bigg]^2
\nonumber\\
&&
\times
\int_{-1}^1\!\! dy \int_{-1}^1\!\! dz 
\,\, V_T (x,y) V_T (y,z) \phi(z,\mu_0)
+O(\alpha_s^3),
\end{eqnarray}
where $\beta_0 = \frac{11}{3} N_c - \frac{2}{3} n_f$.
We can compute $\int_{-1}^{1}dx\, T_0(x)\phi(x,\mu)$ by making use of
Eq.~(\ref{delta-expn}) and the following integrals from 
Ref.~\cite{Jia:2008ep}:
\begin{subequations}
\begin{eqnarray}
f_1(y) &=& 
\int_{-1}^1\! dx \,\, T_0(x) V_T(x,y) 
= \frac{4}{1-y^2}
\left(3 + 2 \log \frac{1-y^2}{4}
\right), 
\\
f_2(z) &=& 
\int_{-1}^1\! dx \int_{-1}^1\! dy 
\,\, T_0(x) V_T(x,y) V_T(y,z)
\nonumber \\ 
&=& 
\frac{4}{1-z^2}
\left[
9+12\log\frac{1-z^2}{4}
+ 4 \left(
\log^2 \frac{1+z}{2}+
\log^2 \frac{1-z}{2}\right)
\right].
\end{eqnarray}
\end{subequations}
The result is
\begin{eqnarray}
\int_{-1}^1 dx \, T_0(x) \phi(x,\mu) 
&=& 4\sum_{k=0}^\infty
\langle x^{2k}\rangle
+
\Bigg[C_F
\frac{\alpha_s(\mu)}{4\pi}\log\frac{\mu^{2}}{\mu_0^2}\Bigg]
\Bigg[
1
+\frac{\beta_0}{2}\frac{\alpha_s(\mu)}{4\pi}
\log\frac{\mu^{2}}{\mu_0^2}
\Bigg]
\nonumber\\
&&\times
\sum_{k=0}^\infty\frac{f_1^{(2k)}(0)}{(2k)!}
\langle x^{2k}\rangle
+
\frac{1}{2}
\Bigg[C_F\frac{\alpha_s(\mu)}{4\pi}
\log\frac{\mu^{2}}{\mu_0^2}\Bigg]^2
\sum_{k=0}^\infty\frac{f_2^{(2k)}(0)}{(2k)!}
\langle x^{2k}\rangle
\nonumber\\
&&
+O(\alpha_s^3),
\end{eqnarray}
where, of course, this expression contains only the leading logarithmic
term in each order in $\alpha_s$. Using 
\begin{subequations}
\begin{eqnarray}
f_1(0)&=&4( 3 - 4 \log 2 ),
\\
f_1^{(2)}(0)&=&8(1 - 4 \log 2),
\\
f_2(0)&=&4(9 - 24 \log 2 + 8\log^2 2),
\\
f_2^{(2)}(0)&=&8(5 - 16 \log 2 + 8\log^2 2),
\end{eqnarray}
\end{subequations}
we obtain
\begin{equation}
\int_{-1}^1 dx \, T_0(x) \phi(x,\mu)=
4c_0(\mu)+4c_2(\mu)\langle x^2\rangle
+O(\langle x^4\rangle), 
\label{LCDA-expn}
\end{equation} 
where
\begin{subequations}
\label{LCDA-expn-coeffs}%
\begin{eqnarray}
c_0(\mu)
&=&1+C_F\Bigg[
\frac{\alpha_s(\mu)}{4\pi}\log\frac{\mu^{2}}{\mu_0^2}\Bigg]
(3 - 4 \log 2)+C_F\Bigg[
\frac{\alpha_s(\mu)}{4\pi}\log\frac{\mu^{2}}{\mu_0^2}\Bigg]^2
\nonumber\\
&&
\times
\Bigg[
C_F\left(\frac{9}{2} - 12 \log 2 + 4\log^2 2\right)
+
\beta_0
\left(\frac{3}{2} - 2 \log 2\right)
\Bigg]+O(\alpha_s^3),
\\
c_2(\mu)&=&
1+C_F
\Bigg[
\frac{\alpha_s(\mu)}{4\pi}\log\frac{\mu^{2}}{\mu_0^2}\Bigg]
(1 - 4 \log 2)
+C_F
\Bigg[
\frac{\alpha_s(\mu)}{4\pi}\log\frac{\mu^{2}}{\mu_0^2}\Bigg]^2
\nonumber\\
&&\times
\Bigg[
C_F
\left(\frac{5}{2} - 8 \log 2 + 4\log^2 2\right) 
+
\beta_0
\left(\frac{1}{2} - 2 \log 2\right)\Bigg]+O(\alpha_s^3)
.
\end{eqnarray}
\end{subequations}
These series converge rapidly. The $\alpha_s^2$ term in $c_2$ is about 
6\% for $\mu_0=m_c$ and about 4\% for $\mu_0=m_b$.

\section{Summary of corrections to the direct amplitude through order 
$\bm{v^2}$}
Now, let us summarize the corrections through order $v^2$ that 
we use in this paper in computing the direct amplitude. 
Our calculations of the
direct amplitude are carried out through order $v^2$ and at leading
order in  $m_Q^2/m_H^2$. They are based on the expression in the second
equality of Eq.~(\ref{phys-amp-lcda}). We expand the LCDA according to
Eq.~(\ref{delta-expn}).

The $\delta(x)$ term in Eq.~(\ref{delta-expn}) was taken into account in
Ref.~\cite{Shifman:1980dk}. There, the coefficient $c_0(\mu)$ in
Eq.~(\ref{LCDA-expn}) was computed to all orders in $\alpha_s$. These
leading logarithms from the evolution of the LCDA were combined with
additional leading logarithms of $m_H^2/m_Q^2$ that arise from the
running of $m_Q$ in the $HQ\bar Q$ coupling\footnote{The logarithms
in $F_{HQ\bar Q}$ are much more important numerically than the
logarithms in $c_0$ because of cancellations that make the coefficients
of the logarithms in $c_0$ small. That is not the case for the
logarithms in $c_2$, which are comparable numerically to the logarithms
in $F_{HQ\bar Q}$.}:
\begin{equation}
F_{HQ\bar Q}(\mu)=[\alpha_s(\mu_0)/\alpha_s(\mu)]^{-3C_F/\beta_0}.
\label{HQbar-evol}
\end{equation}
Finally, the all-orders sums of logarithms were combined with a
fixed-order light-cone calculation of the amplitude through order
$\alpha_s$. The order-$\alpha_s$ logarithm of $m_H^{2}/m_Q^2$ that is
contained in the all-orders sum was subtracted from this fixed-order
calculation in order to avoid double counting. The complete correction
factor for the direct amplitude, relative to the order-$\alpha_s^0$
contribution, is given in Eq.~(78) of Ref.~\cite{Shifman:1980dk}. In
that expression, the LCDA and the $HQ\bar Q$ coupling are evolved from
$2m_Q$ to $m_H$. We evolve from $m_Q$ to $m_H$, instead.\footnote{The
logarithms in the LCDA are collinear logarithms, whose natural cutoff is
$m_Q$. The logarithms in the running mass vanish when $\mu=m_Q$.}
Therefore, we modify the expression in Eq.~(78) of in
Ref.~\cite{Shifman:1980dk} by making the replacement
\begin{equation}
-2\log 2\log\frac{m_H^2}{4m_Q^2}\to -2\log 2\log\frac{m_H^2}{m_Q^2}
\end{equation}
in the last term of that equation. (We have also corrected an obvious
typo: $\log[2(1-\kappa)]\to \log[2(\kappa-1)]$.) We denote this
modified version of the expression in  Eq.~(78) of
Ref.~\cite{Shifman:1980dk} by $g_{SV}$.

For the $\delta^{(2)}(x)$ term in Eq.~(\ref{delta-expn}), we include the
factor $c_2(\mu)$ in Eq.~(\ref{LCDA-expn-coeffs}) and the factor
$F_{HQ\bar Q}(\mu)$ in Eq.~(\ref{HQbar-evol}). These take into account
the leading logarithms of $m_H^2/m_Q^2$ from the evolution of the LCDA
through order $\alpha_s^2$ and the leading logarithms from the running
of the $HQ\bar Q$ coupling to all orders in $\alpha_s$, respectively. 
A fixed-order calculation at order $\alpha_s$ is not available for the
$\delta^{(2)}(x)$ term in Eq.~(\ref{delta-expn}).

The complete expression for the direct amplitude that we use in our 
numerical calculations is then
\begin{equation}
\label{Mdir-calc}
i {\cal M}_{\rm dir}^{\rm calc}
[H\to V+\gamma] =  \sqrt{2 m_V} \phi_0
i {\cal M}_{\rm dir}^{(0)}[H\to V+\gamma]
\bigg[\bigg(1-\frac{5}{6}\langle v^2\rangle \bigg)g_{SV}
+\frac{1}{3}\langle v^2\rangle c_2(\mu)F_{HQ\bar Q}(\mu)\bigg].
\end{equation} 
As we have mentioned, in computing $g_{SV}$, $c_2(\mu)$, and $F_{HQ\bar
Q}$ in this expression, we evolve from  $m_Q$ to $m_H$. When $m_Q=m_b$,
we carry out the evolution with $n_f=5$. When $m_Q=m_c$, we carry out
the evolution in two steps: one from $m_c$ to $m_b$, with $n_f=4$, and
another from $m_b$ to $m_H$, with $n_f=5$.
\section{Decay rate \label{sec:numerical}}
In this section we compute numerical results for the rates for $H\to 
J/\psi +\gamma$ and $H\to \Upsilon +\gamma$. 

First we write the direct 
amplitude in Eq.~(\ref{Mdir-calc}) as
\begin{equation}
{\cal M}_{\rm dir}^{\rm calc}={\cal A}_{\rm dir}
\left(-\epsilon_V^*\cdot \epsilon^*_{\gamma}
+\frac{\epsilon_V^*\cdot p_\gamma  \,
p_V\cdot\epsilon^*_{\gamma}}
{p_\gamma\cdot p_V}\right).
\end{equation}

The indirect amplitude is given by \cite{Bodwin:2013gca}
\begin{equation}
{\cal M}_{\rm ind}={\cal A}_{\rm ind}
\left(-\epsilon_V^*\cdot \epsilon^*_{\gamma}
+\frac{\epsilon_V^*\cdot p_\gamma  \,
p_V\cdot\epsilon^*_{\gamma}}
{p_\gamma\cdot p_V}\right),
\end{equation}
where 
\begin{equation}
{\cal A}_{\rm ind}=\frac{g_{V\gamma}\sqrt{4\pi\alpha(m_V)m_H}}{m_V^2}
\bigg[16\pi\frac{\alpha(m_V)}{\alpha(0)}
\Gamma(H\to\gamma\gamma)\bigg]^{\frac{1}{2}},
\label{Aind}
\end{equation}
and $g_{V\gamma}$ can be written in terms of the width of $V$ into leptons 
\cite{Bodwin:2013gca}:
\begin{equation}
g_{V\gamma}=-
\frac{e_Q}{|e_Q|}\bigg[\frac{3m_V^3\Gamma(V\to 
l^+l^-)}{4\pi \alpha^2(m_V)}\bigg]^{\frac{1}{2}}.
\label{gVgam}
\end{equation}
We remind the reader that $g_{V\gamma}$, 
as computed in Eq.~(\ref{gVgam}), 
already contains all of the corrections of higher 
order in $\alpha_s$ and $v$ that would appear in the NRQCD expression 
for the indirect rate \cite{Bodwin:2006yd,Bodwin:2013gca}.
Note that both ${\cal A}_{\rm dir}$ and ${\cal A}_{\rm ind}$ have
dimensions of mass and are normalized differently than in
Ref.~\cite{Bodwin:2013gca}. We have neglected a small phase in ${\cal
A}_{\rm ind}$ that is about 0.005.
We have dropped terms in Eq.~(\ref{Aind}) that are proportional to
$m_V^2$ divided by combinations of $m_H^2$, $m_t^2$, $m_Z^2$, or
$m_W^2$. The calculation of such terms in Ref.~\cite{Bodwin:2013gca} was
incomplete, in that it did not include the full set of diagrams that is
needed for electroweak gauge invariance.

The sum of the square of the total amplitude over 
the polarizations of the photon and the 
quarkonium is proportional to
\begin{equation}
\label{EE-sum}
\sum_{\rm pol}\left|
-\epsilon_V^*\cdot \epsilon^*_{\gamma}
+\frac{\epsilon_V^*\cdot p_\gamma  \,
p_V\cdot\epsilon^*_{\gamma}}
{p_\gamma\cdot p_V}
\right|^2=2,
\end{equation}
where we have used
\begin{subequations}
\begin{eqnarray}
\label{photon-pol-Feynman}
\sum_{\gamma~{\rm pol}}\epsilon_\gamma^{\mu*} \epsilon_\gamma^{\nu}
&=&-g^{\mu\nu},\\
\sum_{V~{\rm pol}}\epsilon_{V}^{\mu*} \epsilon_{V}^{\nu}
&=&-g^{\mu\nu}+\frac{p_V^\mu p_V^\nu}{p_V^2}.
\end{eqnarray}
\end{subequations}

We then find that the decay rate is 
\begin{equation}
\Gamma(H\to V+\gamma)=2\frac{1}{2m_H}
\frac{m_H^2-m_V^2}{8\pi m_H^2}|{\cal A}_{\rm 
dir} +{\cal A}_{\rm ind}|^2,
\label{rate}
\end{equation}
where the first factor comes from the polarization sum, the second
factor comes from relativistic normalization of the Higgs-boson state,
and the third factor comes from the phase space.

Now let us comment on the choices of scales for the electromagnetic 
coupling $\alpha$. In the direct amplitude, the photon is on shell, and 
so we take $e=\sqrt{4\pi \alpha(0)}$. In the indirect amplitude we use 
$\alpha(m_V)$ to compute $g_{V\gamma}$ from the $V$ leptonic width. We 
also use $e=\sqrt{4\pi \alpha(m_V)}$ for the couplings of the virtual 
photon and $e=\sqrt{4\pi \alpha(0)}$ for the coupling of the real 
photon. We have compensated for the fact that
$\Gamma(H\to\gamma\gamma)$ was computed using $e=\sqrt{4\pi 
\alpha(0)}$. The couplings in the indirect amplitude 
are shown explicitly in Eqs.~(\ref{Aind}) and
(\ref{gVgam}). Note that the dependences on $\alpha(m_V)$ cancel in the 
indirect amplitude.
We use the following value of $\alpha$: 
\begin{equation}
\alpha(0)=1/137.036.
\end{equation}

In evaluating Eq.~(\ref{rate}), we take $m_Q$ to be the pole mass in
order to maintain consistency with the one-loop corrections to the
direct amplitude that we include. We obtain the numerical value of the
pole mass by making use of the one-loop expression that relates the pole
mass to the modified minimal subtraction $(\overline{\textrm{MS}})$
mass.  This procedure has the effect of replacing the pole mass with
the $\overline{\textrm{MS}}$ mass in the expressions through one-loop
order and avoids the issue that the pole mass does not have a definite
value, owing to the presence of an infrared renormalon in its
definition. We use
\begin{subequations}
\begin{eqnarray}
m_c&=&1.483 \pm 0.029~{\rm GeV},\\
m_b&=&4.580 \pm 0.033~{\rm GeV}.
\end{eqnarray}
\end{subequations}
Interpolating the results in 
Ref.~\cite{Bodwin:2007fz} ($J/\psi$) and in
Ref.~\cite{Chung:2010vz} ($\Upsilon$) for the values of $m_Q$ that
we use, we obtain
\begin{subequations}
\begin{eqnarray}
\phi_0^2(J/\psi)&=&\phantom{-}0.0729\pm 0.0109~{\rm GeV}^3,\\
\langle v^2\rangle(J/\psi)&=&\phantom{-}0.201\pm 0.064,\\
\phi_0^2[\Upsilon(1S)]&=&\phantom{-}0.512\pm 0.035~{\rm GeV}^3,\\
\langle v^2\rangle[\Upsilon(1S)]&=& -0.00920\pm 0.00348,\\ 
\phi_0^2[\Upsilon(2S)]&=&\phantom{-}0.271\pm 0.019~{\rm GeV}^3,\\
\langle v^2\rangle[\Upsilon(2S)]&=&\phantom{-} 0.0905\pm 0.0100,\\ 
\phi_0^2[\Upsilon(3S)]&=&\phantom{-}0.213\pm 0.015~{\rm GeV}^3,\\
\langle v^2\rangle[\Upsilon(3S)]&=& \phantom{-}0.157\pm 0.017.
\end{eqnarray}
\end{subequations}

We take $m_H=125.9\pm 0.4$~GeV, and we obtain $\Gamma(H\to
\gamma\gamma)=9.565\times 10^{-6}$~GeV from the values of the
Higgs-boson total width and branching fraction to $\gamma\gamma$ in
Refs.~\cite{Dittmaier:2011ti,Dittmaier:2012vm}.

We estimate the uncertainties in the indirect amplitude along the lines
that were suggested in footnote~2 of Ref.~\cite{Bodwin:2013gca}. In
$\Gamma(H\to \gamma \gamma)$, we take the uncertainty from uncalculated
higher-order corrections to be 1\%, and the uncertainties that arise from
the uncertainties in the top-quark mass $m_t$ and the $W$-boson mass
$m_W$ to be 0.022\% and 0.024\%, respectively. We take the uncertainties
in the leptonic decay widths to be 2.5\% for the $J/\psi$ and 1.3\% for
the $\Upsilon$. We estimate the uncertainties in the indirect amplitude
from uncalculated mass corrections to be $m_V^2/m_H^2$. We have not
included the effects of the uncertainty in $m_H$, as it is expected that
that uncertainty will be significantly reduced in Run~II of the LHC.

The uncertainties in the direct amplitude arise primarily from the
uncertainties in $\phi_0$, $\langle v^2\rangle$, and uncalculated
corrections of order $\alpha_s^2$, order $\alpha_s v^2$, and order $v^{4}$. We
estimate the order-$\alpha_s^2$ correction to be $2\%$, the
order-$\alpha_s v^2$ correction to be $5\%$ for the $J/\psi$ and
$1.5\%$ for the $\Upsilon$, and the order-$v^4$ correction to be $9\%$
for the $J/\psi$ and $1\%$ for the $\Upsilon$. The uncertainties in
the direct amplitude that arise from the uncertainties in $m_c$ and 
$m_b$ are 0.6\% in the case of the $J/\psi$ and 0.1\% in the
case of the $\Upsilon$, and so they are negligible in comparison with the
other uncertainties in the direct amplitude.

Our results for the widths are\footnote{
We do not include results for the $\psi(2S)$ because a value for
$\langle v^2\rangle[\psi(2S)]$ does not exist in the literature and because 
it is likely that $v^2$ for the $\psi(2S)$ is so large that the theoretical
uncertainties in the width would be very large.} 
\begin{subequations}                                              
\label{widths1}
\begin{eqnarray}
\Gamma(H\to J/\psi+\gamma)&=& 
\big| (11.9 \pm 0.2)   - 
   (1.04 \pm 0.14) \kappa_{c} \big|^2
\times 10^{-10}~{\rm GeV},
\\
\Gamma[H\to \Upsilon(1S)+\gamma]&=&
\big| (3.33 \pm 0.03) -
 (3.49 \pm 0.15)\kappa_{b} \big|^2 
\times 10^{-10}~{\rm GeV},
\\
\Gamma[H\to \Upsilon(2S)+\gamma]&=&
\big| (2.18 \pm 0.03) -
 (2.48 \pm 0.11)\kappa_{b} \big|^2 
\times 10^{-10}~{\rm GeV},
\\
\Gamma[H\to \Upsilon(3S)+\gamma]&=&
\big| (1.83 \pm 0.02) -
 (2.15 \pm 0.10)\kappa_{b} \big|^2 
\times 10^{-10}~{\rm GeV}.
\end{eqnarray}                                                          
\end{subequations}
The SM values for the widths ($\kappa_{Q}=1$) are
\begin{subequations}                      
\label{widths2}
\begin{eqnarray}              
\Gamma_{\rm SM}(H\to J/\psi+\gamma)&=& 
1.17 ^{+0.05}_{-0.05} \times 10^{-8} ~{\rm GeV},\\
\Gamma_{\rm SM}[H\to \Upsilon(1S)+\gamma]&=& 
2.56 ^{+7.30}_{-2.56} \times 10^{-12} ~{\rm GeV},\\
\Gamma_{\rm SM}[H\to \Upsilon(2S)+\gamma]&=& 
8.46 ^{+7.79}_{-5.35} \times 10^{-12} ~{\rm GeV},\\
\Gamma_{\rm SM}[H\to \Upsilon(3S)+\gamma]&=& 
10.25 ^{+7.33}_{-5.45} \times 10^{-12} ~{\rm GeV}.
\end{eqnarray}                    
\end{subequations}
Using $\Gamma(H)=4.195_{-0.159}^{+0.164}\times 10^{-3}$~GeV~\cite{LHCXS}, 
we obtain the following results for the branching
fractions in the SM:
\begin{subequations}                      
\label{branchingratios2}
\begin{eqnarray}              
\mathcal{B}_{\rm SM}(H\to J/\psi+\gamma)&=& 
2.79_{-0.15}^{+0.16} \times 10^{-6} ,\\
\mathcal{B}_{\rm SM}[H\to \Upsilon(1S)+\gamma]&=& 
6.11_{-6.11}^{+17.41} \times 10^{-10} ,\\
\mathcal{B}_{\rm SM}[H\to \Upsilon(2S)+\gamma]&=& 
2.02_{-1.28}^{+1.86} \times 10^{-9} ,\\
\mathcal{B}_{\rm SM}[H\to \Upsilon(3S)+\gamma]&=& 
2.44_{-1.30}^{+1.75} \times 10^{-9} .
\end{eqnarray}                    
\end{subequations}

In comparison with the results in Ref.~\cite{Bodwin:2013gca}, the
coefficient of $\kappa_{c}$ has been reduced by about 30\%, and
the coefficient of $\kappa_{b}$ has been reduced by about
12\%. In the case of the coefficient of $\kappa_{c}$, the reduction 
arises as follows:  a reduction of 11\% from including the relativistic
corrections; a reduction of 18\% from summing logarithms by evolving
from the scale $m_c$ rather than from the scale $2m_c$ and from using
a variable flavor number rather than a fixed flavor number $n_f=3$; and a
reduction of 3\% from using $\alpha(0)$ rather than $\alpha(m_H/2)$ for
the electromagnetic coupling of the on-shell quark. In the case of the
coefficient of $\kappa_{b}$, the reduction arises as follows:  a
reduction of 0\% from the relativistic corrections; a reduction of 9\%
from summing logarithms by evolving from the scale $m_b$ rather than
from the scale $2m_b$, and from using $n_f=5$ rather than $n_f=3$; and a
reduction of 3\% from using $\alpha(0)$ rather than $\alpha(m_H/2)$ for
the electromagnetic coupling of the on-shell quark. In addition, there
are changes in the coefficients of $\kappa_c$ and $\kappa_b$ of less than
1\% that come from changes in the values of $m_c$, $m_b$, and $m_H$.
\section{Summary and discussion\label{sec:summary}}
In this paper, we have calculated relativistic corrections to the direct
decay amplitude that appears in the Higgs-boson width $\Gamma(H\to
V+\gamma)$, where $V$ is a $J/\psi$ or an $\Upsilon(nS)$ state with 
$n=1,2,3$.

Using NRQCD factorization methods, we have calculated corrections to all
orders in the heavy-quark velocity $v$ for NRQCD LDMEs of the form in
Eq.~(\ref{v2n-def}), keeping the exact dependence on the ratio of the
heavy-quark mass $m_Q$ to the Higgs-boson mass $m_H$. The  result of
this calculation is given in Eq.~(\ref{phys-amp}), where $R(v^2)$ is
given in Eq.~(\ref{R-v2}).

Using light-cone methods, we have calculated relativistic corrections
through order $v^2$ at the leading order in $m_Q^2/m_H^2$. In the
light-cone method, the corrections in order $v^2$ arise from both the
$x^0$ term and the $x^2$ term in the hard-scattering kernel $T_0(x)$ 
[Eq.~(\ref{eq:hard_kernel})],
where $x$ is the light-cone momentum fraction. In the case of the
corrections that arise from the $x^0$ term, we have applied existing
corrections of order $\alpha_s$ and corrections from a summation of
leading logarithms of $m_H^2/m_Q^2$ to all orders in $\alpha_s$
\cite{Shifman:1980dk}. In the case of the corrections that arise from
the $x^2$ term, we have computed and applied corrections from leading
logarithms of $m_H^2/m_Q^2$. 
We have computed leading logarithms from the running of the $HQ\bar Q$ coupling
to all orders in $\alpha_s$ and leading logarithms from the evolution of the
LCDA through order $\alpha_s^2$.
Leading
logarithmic corrections of order $\alpha_s^3$ and higher are estimated to
contribute at the level of about $1\%$. The complete result from
applying these various corrections is given in Eq.~(\ref{Mdir-calc}). We
used this result in our numerical calculations.

Our numerical results for the widths $\Gamma(H\to J/\psi +\gamma)$ and
$\Gamma(H\to \Upsilon(nS) +\gamma)$ are given in
Eqs.~(\ref{widths1}) and (\ref{widths2}), where $\kappa_{Q}$ in
Eq.~(\ref{widths1}) parametrizes the deviation of the $HQ\bar Q$
coupling from the SM value. In comparison with the results in
Ref.~\cite{Bodwin:2013gca}, the coefficient of $\kappa_{c}$ has been
reduced by about 30\%, and the coefficient of $\kappa_{b}$ in
$\Gamma(H\to \Upsilon(1S) +\gamma)$ has been reduced by about 12\%.
The relativistic corrections themselves contribute only about 11\% and
0\% of this reduction, respectively. The bulk of the reduction comes
from the use of a different procedure for summing leading logarithms of
$m_H^2/m_Q^2$, namely, evolving from the scale $m_Q$ rather than from
the scale $2m_Q$, and from the use of a variable flavor number, rather than
$n_f=3$. The relativistic corrections are very small in the
$\Upsilon(1S)$ case, owing to a cancellation in the corresponding
dimensionally regulated NRQCD LDME that makes $\langle v^2\rangle$
anomalously small. We note that, for SM couplings, the destructive
interference between the direct and indirect amplitudes is less complete
in the $\Upsilon(2S)$ and $\Upsilon(3S)$ channels than in the
$\Upsilon(1S)$ channel, and, hence, the SM rates are larger in the former
channels.

More significant than the changes in the values of the coefficients of
$\kappa_{Q}$ in Eq.~(\ref{widths1}) 
are the changes in the theoretical uncertainties for those
coefficients. Relative to the uncertainties that were given in
Ref.~\cite{Bodwin:2013gca}, they have been reduced by about a factor of
3.3 for the coefficient of $\kappa_c$ in $\Gamma(H\to J/\psi +\gamma)$
and by about a factor of 2.8 for
the coefficient of $\kappa_{b}$ in $\Gamma(H\to \Upsilon(1S)
+\gamma)$.

In the case of the channel $H\to J/\psi +\gamma$, our values for the
decay rate indicate that it should be possible to collect a sample of
about 50 events in a high-luminosity run at the LHC~\cite{Bodwin:2013gca}. 
This would imply a statistical error in the
measurement of $\Gamma(H\to J/\psi +\gamma)$ of 14\% and a statistical
error in the determination of $\kappa_{c}$ of about 40\%. The
latter error is comparable to the theoretical uncertainty in the
coefficient of $\kappa_{c}$ that existed in the absence of a
calculation of relativistic corrections. The inclusion of the
relativistic corrections that we have calculated reduces that
uncertainty to about 16\% and opens the door to determinations of the
$Hc\bar c$ coupling at higher levels of precision.
\begin{acknowledgments}
We thank Eric Braaten and Kostas Nikolopoulos for helpful discussions.
The work of G.T.B.\ and H.S.C.\ is supported by the U.S.\ Department of
Energy, Division of High Energy Physics, under contract
DE-AC02-06CH11357. The work of J.-H.E.\ is supported by Global Ph.D.
Fellowship Program through the National Research Foundation of
Korea (NRF) funded by the Ministry of Education (Grant No.
NRF-2012H1A2A1003138). 
The work of F.P. is supported by the U.S.\ Department of
Energy, Division of High Energy Physics, under contract DE-FG02-91ER40684 and the grant DE-AC02-06CH11357.  The submitted manuscript has been created in part
by UChicago Argonne, LLC, Operator of Argonne National Laboratory.
Argonne, a U.S.\ Department of Energy Office of Science laboratory, is
operated under Contract No. DE-AC02-06CH11357. The U.S. Government
retains for itself, and others acting on its behalf, a paid-up
nonexclusive, irrevocable worldwide license in said article to
reproduce, prepare derivative works, distribute copies to the public,
and perform publicly and display publicly, by or on behalf of the
Government.
\end{acknowledgments}
\appendix
\section{Eigenfunction evolution\label{sec:eigen-evol}}
In this appendix, we solve the LCDA evolution equation 
(\ref{LCDA-evol}) in terms of the eigenfunctions of the evolution kernel $V_T$.

The kernel $V_T (x,y)$ has eigenfunctions 
\begin{equation}
G_n(x) = \frac{1-x^2}{4}
C_n^{3/2} (x), 
\end{equation}
where $C_n^{3/2} (x)$ is a Gegenbauer polynomial. The
eigenfunctions satisfy\footnote{See, for example, 
Ref.~\cite{Jia:2008ep}.}
\begin{equation}
\frac{1}{2} \int_{-1}^1 dy V_T (x,y)
G_n(y) = - \gamma_n G_n(x),
\end{equation}
where the eigenvalues $\gamma_n$ are given by 
\begin{equation}
\gamma_n = \frac{1}{2} + 2 \sum_{j=2}^{n+1} \frac{1}{j}.
\end{equation}
Following Ref.~\cite{Jia:2008ep}, we find a formal solution by 
writing
\begin{equation}
\phi(x,\mu) = 
\sum_n \phi_n (\mu) G_n( x ),
\label{eigen-expn}
\end{equation}
where the $\phi_n (\mu)$ can be found by using the orthogonality of the
Gegenbauer polynomials:
\begin{equation}
\phi_n (\mu) = 
\frac{2 (2 n+3)}{(n+1) (n+2)} \int_{-1}^1 dx\, C_n^{3/2} (x) 
\phi(x,\mu). 
\label{orthog}
\end{equation}
The amplitude $i {\cal M}$ is proportional to $\int_{-1}^1 dx\, 
T_0 (x) \phi(x,\mu)$. Using Eq.~(\ref{eigen-expn}), we can 
write
\begin{eqnarray}
\label{eq:appen-imathcalm}
&&\int_{-1}^1 dx\, T_0 (x) \phi(x,\mu) = 
\sum_{n=0}^{\infty}
 \phi_{n} (\mu) \int_{-1}^1 dx\, C_n^{3/2} (x) 
\nonumber \\ 
&&= 
2\sum_{n=0}^{\infty}\phi_{2n} (\mu)
= 2\sum_{n=0}^{\infty}\phi_{2n}(\mu_0 )
\left[ \frac{\alpha_s(\mu)}{\alpha_s (\mu_0)} \right]^{d_{2n}}, 
\end{eqnarray}
where we have used the facts that 
$\int_{-1}^1 dx \, C_{2n}^{3/2} (x) = 2$ and
$\int_{-1}^1 dx \, C_{2n+1}^{3/2} (x) = 0$ for $n$ a non-negative 
integer, and we have defined $d_{2n}\equiv 2 C_F \gamma_{2n}/\beta_0$, 
with $\beta_0 = \frac{11}{3} N_c - \frac{2}{3} n_f$.

In order to find the coefficients $\phi_{2n}(\mu_0)$ from 
Eq.~(\ref{orthog}), 
we expand $\phi(x,\mu_0)$ formally, using 
Eq.~(\ref{delta-expn}).
For $n$ a non-negative integer, we have 
\label{eq:appen-phi2nmu0}
\begin{eqnarray}
\phi_{2n} (\mu_0) 
&=& 
\frac{2 (4 n+3)}{(2n+1) (2n+2)} 
\sum_{k=0}^\infty
\frac{\langle x^{2k} \rangle}{(2k)!}
\frac{d^{2k}}{dx^{2k}} C_{2n}^{3/2} (0)
\nonumber\\
&=& 
\frac{2 (4 n+3)}{(2n+1) (2n+2)} 
\sum_{k=0}^\infty
\frac{\langle x^{2k} \rangle}{(2k)!}
(4k+1)!! C_{2(n-k)}^{(4k+3)/2} (0)
\nonumber \\
&=&
\frac{2 (4 n+3)}{(2n+1)(2n+2)} 
\sum_{k=0}^\infty
(-1)^{n-k}\langle x^{2k} \rangle
\frac{(2n+2k+1)!!}
{(2k)!(2n-2k)!!}. \phantom{xxx}
\end{eqnarray}
Here, we have used the recurrence relation
\begin{equation}
\frac{d}{dx} C_n^{\lambda/2} (x) = \lambda C_{n-1}^{(\lambda+2)/2} (x)
\end{equation}
and the values of the Gegenbauer polynomials at zero argument, 
\begin{subequations}
\begin{eqnarray}
C_{2n+1}^{\lambda/2} (0) 
&=&0,\\
C_{2n}^{\lambda/2} (0)&=&
\frac{(-1)^n\Gamma(n+\frac{\lambda}{2})}{n!\Gamma(\frac{\lambda}{2})}
=
\frac{(-1)^n}{(2n)!!}
\frac{(\lambda+2n-2)!!}{(\lambda-2)!!}.
\end{eqnarray}
\end{subequations}

Taking into account the effect of the running of the $HQ\bar{Q}$
coupling, (that is, the running of the quark mass), whose anomalous
dimension is $-3 C_F$, we can write
\begin{equation}
\int_{-1}^1 dx\, T_0(x) \phi(x,\mu)=
4\sum_{k=0}^\infty
c_{2k}(\mu) \langle x^{2k}\rangle,
\end{equation}
where 
\begin{equation}
c_{2k}(\mu)
=
\sum_{n=0}^\infty
\frac{(-1)^{n-k}(4 n+3)}{(2n+1)(2n+2)} 
\frac{(2n+2k+1)!!}
{(2k)!(2n-2k)!!}
\left[ \frac{\alpha_s(\mu)}{\alpha_s (\mu_0)} \right]^{d_{2n}+3 C_F/\beta_0}
\end{equation}
contains all of the leading logarithms of $m_H^2/m_Q^2$. The expression
for $c_0(\mu)$ reproduces the expression in Eq.~(58) of
Ref.~\cite{Shifman:1980dk}.

Note that, for large $n$, the $n$th term of $c_{2k}(\mu)$ is equal to
\begin{eqnarray}
&&
(-1)^{n-k}
n^{2k-1}
\frac{(2n+1)!!}{(2n)!!}
\frac{2^{2k}}
{(2k)!}
\left[ \frac{\alpha_s(\mu)}{\alpha_s (\mu_0)} \right]^{d_{2n}+3 C_F/\beta_0}
\nonumber\\
&\sim& 
(-1)^{n-k}
\frac{2^{2k+1}}{\sqrt{\pi}(2k)!}
n^{2k-1/2}
\left[ \frac{\alpha_s(\mu)}{\alpha_s (\mu_0)} 
\right]^{(4C_F/\beta_0)(\gamma_{{}_{\rm E}}+\log2)}
n^{(4C_F/\beta_0)\log[\alpha_s(\mu)/\alpha_s (\mu_0)]}.
\end{eqnarray}
Hence, the series for $c_{2k}(\mu)$ converges if and only if
\begin{equation}
\frac{4 C_F}{\beta_0}\log
\frac{\alpha_s(\mu)}{\alpha_s(\mu_0)}<-2k+\frac{1}{2}.
\end{equation}
For $k=0$, this convergence condition is satisfied for $\mu=m_H$ and 
$\mu_0=m_c$ or $\mu_0=m_b$. However, for $k\geq 1$, it is not satisfied
for $\mu=m_H$ and $\mu_0=m_c$ or $\mu_0=m_b$.


\end{document}